# In-situ wavelength calibration without a calibration target: use of Fraunhofer lines after landing on planetary surface

Mori, Shoki, Department of Earth and Planetary Science, The University of Tokyo, 7-3-1 Hongo, Bunkyo, Tokyo 113-0033, Japan, smori@eps.s.u-tokyo.ac.jp
Böttger, Ute, Deutsches Zentrum für Luft- und Raumfahrt e.V. (DLR), Institute of Optical Sensor Systems, 12489 Berlin, Rutherfordstr. 2, Germany, Ute.Boettger@dlr.de,
Buder, Maximilian, Deutsches Zentrum für Luft- und Raumfahrt e.V. (DLR), Institute of Optical Sensor Systems, 12489 Berlin, Rutherfordstr. 2, Germany, Maximilian.Buder@dlr.de;
Cho, Yuichiro, Department of Earth and Planetary Science, The University of Tokyo, 7-3-1 Hongo, Bunkyo, Tokyo 113-0033, Japan, cho@eps.s.u-tokyo.ac.jp
Dietz, Enrico, Deutsches Zentrum für Luft- und Raumfahrt e.V. (DLR), Institute of Optical Sensor Systems, 12489 Berlin, Rutherfordstr. 2, Germany, Enrico.Dietz@dlr.de;
Hagelschuer, Till, Deutsches Zentrum für Luft- und Raumfahrt e.V. (DLR), Institute of Optical Sensor Systems, 12489 Berlin, Rutherfordstr. 2, Germany, Till.Hagelschuer@dlr.de;
Hübers, Heinz-Wilhelm, Deutsches Zentrum für Luft- und Raumfahrt e.V. (DLR), Institute of Optical Sensor Systems, 12489 Berlin, Rutherfordstr. 2, Germany, Heinz-Wilhelm.Huebers@dlr.de;
Kameda, Shingo, Department of Physics, College of Science, Rikkyo University, 3-34-1 Nishiikebukuro, Toshima, Tokyo 171-8501, Japan, kameda@rikkyo.ac.jp
Kopp, Emanuel, Deutsches Zentrum für Luft- und Raumfahrt e.V. (DLR), Institute of Optical Sensor Systems, 12489 Berlin, Rutherfordstr. 2, Germany, Emanuel.Kopp@dlr.de;
Prieto-Ballesteros, Olga, Centro de Astrobiología (CAB-INTA-CSIC), Ctra. Ajalvir, Km 4, 28850 Torrejón de Ardoz, Spain, prietobo@inta.es;
Rull, Fernando*, Universidad de Valladolid – GIR ERICA, Av. Francisco valles, 8, Parque Tecnólogico de Boecillo, Parcela 203, E-47151 Boecillo, Valladolid, Spain, rull@fmc.uva.es,
Ryan, Conor, Deutsches Zentrum für Luft- und Raumfahrt e.V. (DLR), Institute of Optical Sensor Systems, 12489 Berlin, Rutherfordstr. 2, Germany, Conor.Ryan@dlr.de;
Schröder, Susanne, Deutsches Zentrum für Luft- und Raumfahrt e.V. (DLR), Institute of Optical Sensor Systems, 12489 Berlin, Rutherfordstr. 2, Germany, Susanne.Schroeder@dlr.de;
Sugita, Seiji, Department of Earth and Planetary Science, The University of Tokyo, 7-3-1 Hongo, Bunkyo, Tokyo 113-0033, Japan, sugita@eps.s.u-tokyo.ac.jp
Tabata, Haruhisa, Department of Earth and Planetary Science, The University of Tokyo, 7-3-1 Hongo, Bunkyo, Tokyo 113-0033, Japan, tabata@eps.s.u-tokyo.ac.jp
Usui, Tomohiro, Japan Aerospace Exploration Agency (JAXA), Institute of Space and Astronautical Science, Department of Solar System Sciences, 3-1-1 Yoshinodai, Chuo, Sagamihara, Kanagawa, 252-5210, Japan, usui.tomohiro@jaxa.jp;
Yumoto, Koki, Department of Earth and Planetary Science, The University of Tokyo, 7-3-1 Hongo, Bunkyo, Tokyo 113-0033, Japan, yumoto@eps.s.u-tokyo.ac.jp

## Abstract

Accurate wavelength calibration is critical for qualitative and quantitative spectroscopic measurements. Many spectrometers for planetary exploration are equipped with onboard calibration sources. However, such calibration sources are not always available because planetary lander missions often have strong limitation in size and mass. In this study, we propose and validate a wavelength calibration method using solar Fraunhofer lines observed in reflective spectra. As a result, for a visible Raman spectrometer, the accuracy is better than 0.6 cm-1 in 0-4000 cm-1 range, and the magnesium abundance of olivine is estimated more accurately than 2%.

## 1. Introduction

Spectroscopy is a strong tool for planetary exploration. For example, infrared spectroscopy enables the estimation of surface minerals because it contains information on material vibrational characteristics. Spectroscopy also informs gas composition of atmospheres. Thus, many planetary exploration missions have been equipped with various spectrometers, and JAXA's Martian Moons eXploration Mission (MMX) is no exception.

Spectroscopy requires calibration. In particular, quantitative analyses require very accurate calibration. Thus, many spectrometers for planetary exploration are equipped with reference lamps on board. For example, The Visible and Infrared Thermal Imaging Spectrometer-M on board Venus Express has two reference lamps (Cardesin Moinelo et al., 2010; Melchiorri et al., 2003). ChemCam, the combined instrument of laser-induced breakdown spectroscopy and camera onboard the Curiosity rover, has ten calibration targets such as titanium plate in it (Wiens et al., 2012).

On the other hand, planetary exploration missions often have strong payload limitations. Thus, many science instruments have simple structures, and sometimes calibration targets are omitted. For example, MMX has a small rover, and it also has a strong payload limitation. The rover has a small Raman spectrometer called RAX, standing for a Raman spectrometer for MMX. RAX is a very small and lightweight Raman spectrometer without onboard calibration target (Bertrand et al., 2019; Cho et al., 2021; Hagelschuer et al., 2019; Schröder et al., 2020).

Quantitative analysis with Raman spectroscopy requires high precision and accuracy of wavenumber calibration. Estimation of magnesium abundance of olivine with the accuracy of 1% needs measurement of the olivine doublet at 800 $cm^{-1}$ with the accuracy of 0.1 $cm^{-1}$ (Kuebler et al., 2006). Thus, to maximize the capability of small and lightweight spectrometers, a calibration method without a calibration target is needed. The Spectral Profiler of the Selenological and Engineering Explorer was calibrated without calibration targets by using the wavelength-dependent wavy pattern, reflecting the transmission of the low pass filter (Yamamoto et al., 2011). In this study, we propose and validate another method for wavelength calibration with no calibration targets by utilization of Fraunhofer lines observed in reflective spectra.

## 2. Method

### 2.1. Instruments and settings

RAX has the breadboard model (BBM) in German Aerospace Center and the University of Tokyo. The BBM is designed to acquire the same spectra as the flight model (Cho et al., 2021; Rodd-Routley et al., 2021). To simulate the albedo and flat spectrum of the surface of Phobos, a 1% Spectralon reflective standard was placed under the objective lens as a target (Figure 1). The BBM was moved outdoors for sunlight. For the reference wavelength calibration, a conventional neon lamp was used. Also, a halogen lamp (Ocean Optics LS-1-CAL) was used for sensitivity calibration of the detector.

Gain was set to 0 dB in order to reduce noise, and the exposure time was determined by taking several snapshots with various exposure times.

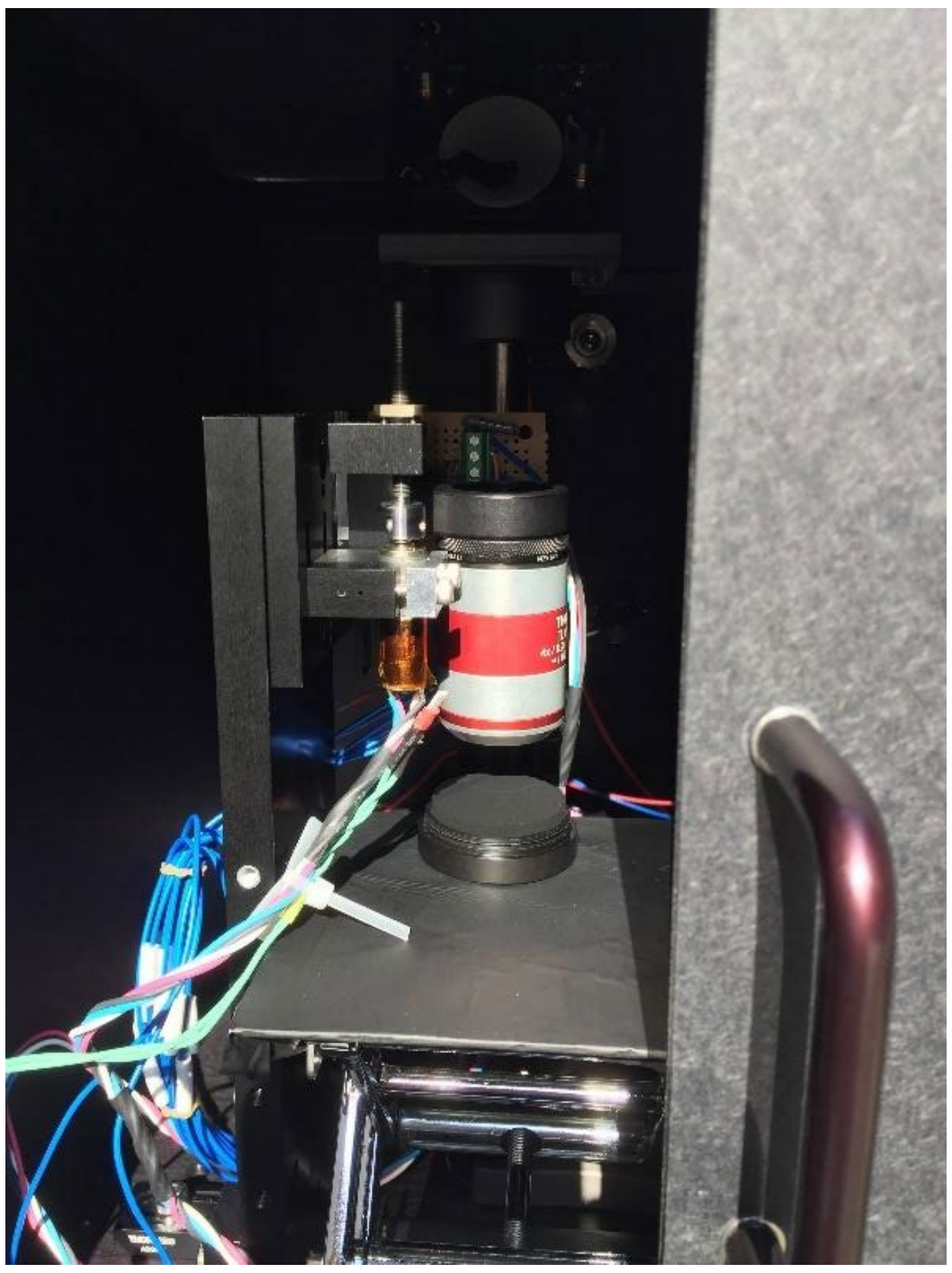

**Figure 1.** Setting of the experiment. Sunlight was reflected by the 1% reflective standard placed under the objective lens.

### 2.2. Fraunhofer lines determination

First, reference wavelength calibration was executed with the spectrum of a neon lamp. The peak pixel number was matched to the reference Ne spectral lines (*Strong Lines of Neon ( Ne )*, n.d.), and the wavelength for each detector pixel number was estimated by fitting a third polynomial function. Next, to reduce the effect of the optical response function of the BBM, sensitivity calibration was done by LS-1-CAL. The sensitivity factor was calculated by dividing the reference radiance by the detector readout values.

After these calibrations, the acquired spectrum was matched to the literature values (Moore et al., 1966) for validation. The quality of the acquired solar spectrum is determined by the difference between the detected peak wavelength and the literature value. The regions where the absorption of Earth's atmosphere is major (e.g., 680- nm) were ignored.

### 2.3. Superimposed solar spectrum of BBM

Because the solar spectrum has quite a lot of absorptive lines, some strong lines are overlapped with adjacent lines. With these lines, the observed peaks do not exactly match the literature values. To correct this effect, the superimposed spectrum was calculated from the literature values.

First, the peak width of a single bright line of the BBM was determined by fitting a gauss function to each line of the neon spectrum. The superimposed spectrum was calculated by summing up the gauss functions for each absorptive line. The width of the gauss function was the device line width, and the height of that was the equivalent width for each line. The superimposed spectrum was calculated for the peak where the largest difference derived in the former section. The calculation was repeated until every difference became small enough. In this study, the threshold was set to 1.33 times for each pixel size. Ideally, the difference should be smaller than the pixel size, but this value was determined by taking some margin, considering device thermal changes and the effect of Earth's atmosphere.

### 2.4. Calibration with Fraunhofer lines

After this preprocessing, the observed dip pixel number and the superimposed dip wavelength are gained. Using this set, the wavelength for each pixel was also calculated by fitting a third polynomial function.

The accuracy was shown when the difference from the reference calibration in Raman shift. Wavelength is converted to Raman shift with the following equation:

$$x = 10^7 \left(\frac{1}{\lambda_c} - \frac{1}{\lambda}\right),$$

where $x$ is Raman shift in $cm^{-1}$, $\lambda$ is the wavelength of each pixel in nm, and $\lambda_c$ is the excitation light wavelength. The BBM of RAX has an excitation laser with a wavelength of 532.04 nm.

## 3. Results and discussions

On August 27, 2021, we acquired a solar spectrum from 15 o'clock to 15:30. After it, we acquired spectra of neon and halogen lamp. It was conducted in the University of Tokyo. Figure 2 is the solar spectrum acquired by the BBM. Various absorption lines were observed.

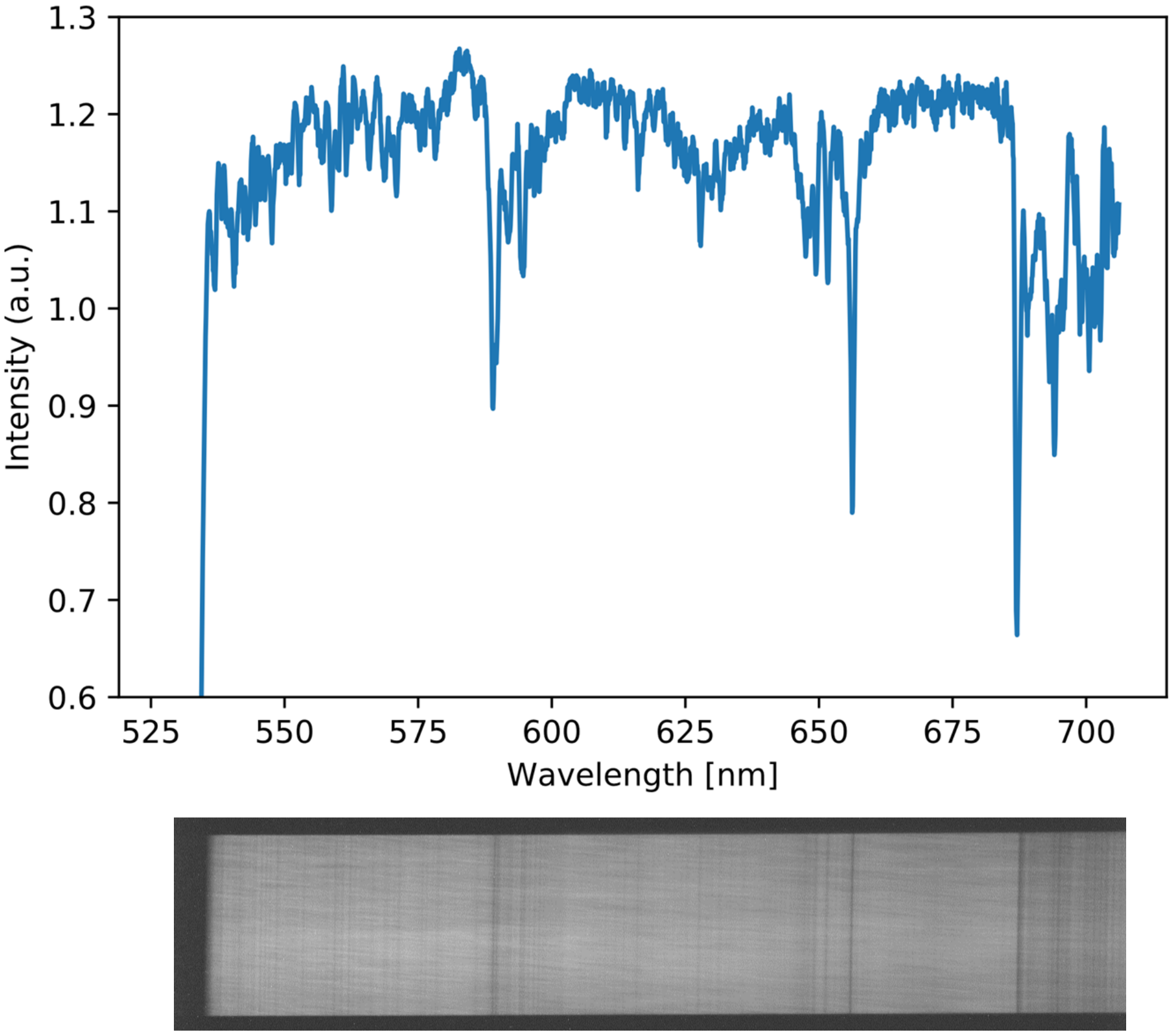

**Figure 2.** Solar spectrum acquired by the BBM.

With the neon lamp, Figure 3 shows wavelength was estimated to be

$$\lambda(x) = a + bx + cx^2 + dx^3$$
$$a = 5.27920603 \times 10^2\ nm$$
$$b = 8.88997848 \times 10^{-2}\ nm\ pix^{-1}$$
$$c = 2.06930744 \times 10^{-7}\ nm\ pix^{-2}$$
$$d = -5.38926957 \times 10^{-10}\ nm\ pix^{-3},$$

where $x$ is the pixel number. This means the size of the pixel is approximately 0.09 nm. The residual error ranged from -0.05 nm to 0.04 nm, comparable to the pixel size limit. The residual sum of squares was 0.01677. The FWHM of the BBM was estimated to be 0.50 nm.

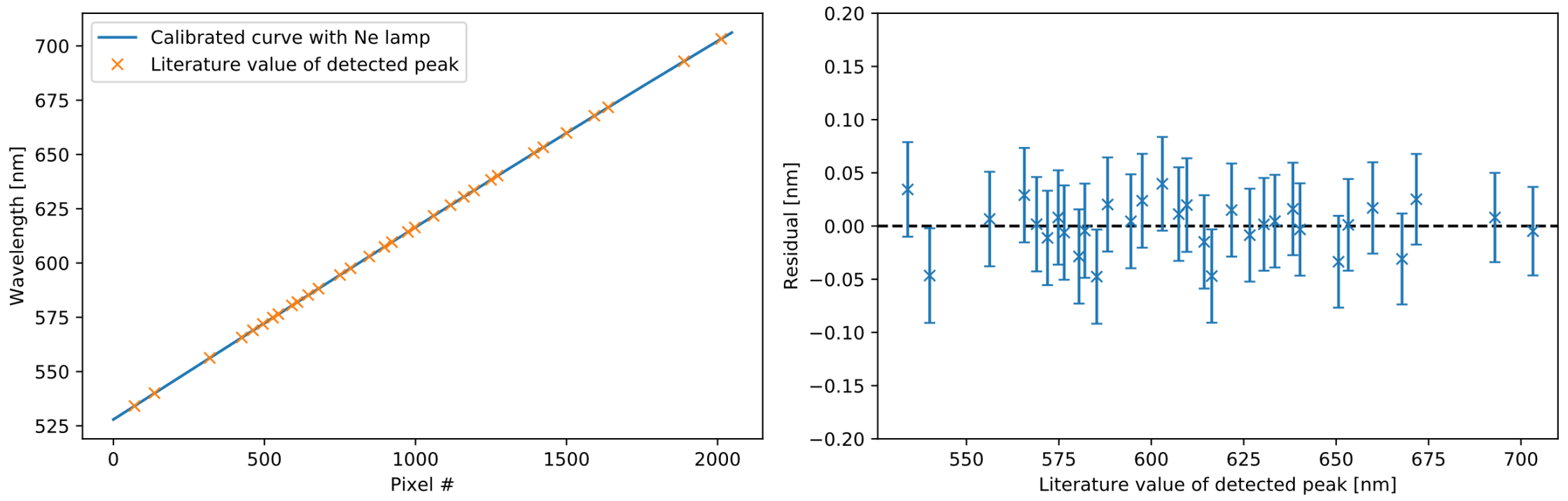

**Figure 3.** Result of the calibration with Ne lamp. Left: the calibrated curve and the detected peaks. Right: Residual of each detected peak.

Figure 2 shows the solar spectrum after wavelength and sensitivity calibration. The absorption of sodium and hydrogen were prominent at 589 nm and 656 nm, respectively. Figure 4 shows an example of matching literature values and detected peaks. In this manner, 29 absorption lines in table 1 are extracted. With these lines, another wavelength calibration was done in figure 6. The parameters were

$$a = 5.27942764 \times 10^{2}\ nm$$
$$b = 8.87548489 \times 10^{-2}\ nm\ pix^{-1}$$
$$c = 4.46774974 \times 10^{-7}\ nm\ pix^{-2}$$
$$d = -6.40485752 \times 10^{-10}\ nm\ pix^{-3}$$

The residual error ranged from -0.06 nm to 0.06 nm, and the residual sum of square was 0.03295. This is slightly larger than the calibration with Ne lamp because the effect of the Earth's atmosphere was not completely removed. Comparing calculated Raman shifts derived from Ne lamp calibration and Fraunhofer lines calibration, Figure 6 shows the difference in 0-4000 $cm^{-1}$ range was less than $\pm 0.5\ cm^{-1}$. In addition, between 250 $cm^{-1}$ to 2000 $cm^{-1}$, the difference was smaller than $\pm 0.2\ cm^{-1}$. This means the accuracy in the most important region is good enough. Using this difference, the olivine spectrum of simulated calibration with Fraunhofer lines was calculated in Figure 7. The original spectrum was acquired with BBM on October 30 and calibrated with a neon lamp. Adding the difference derived in Figure 6 to the Raman shift axis of the original spectrum, the simulated spectrum was generated. Figure 7 shows that the differences in the olivine doublet's peak positions are less than 0.2 $cm^{-1}$. This means that the forsterite number of olivine is determined more accurately than 1%.

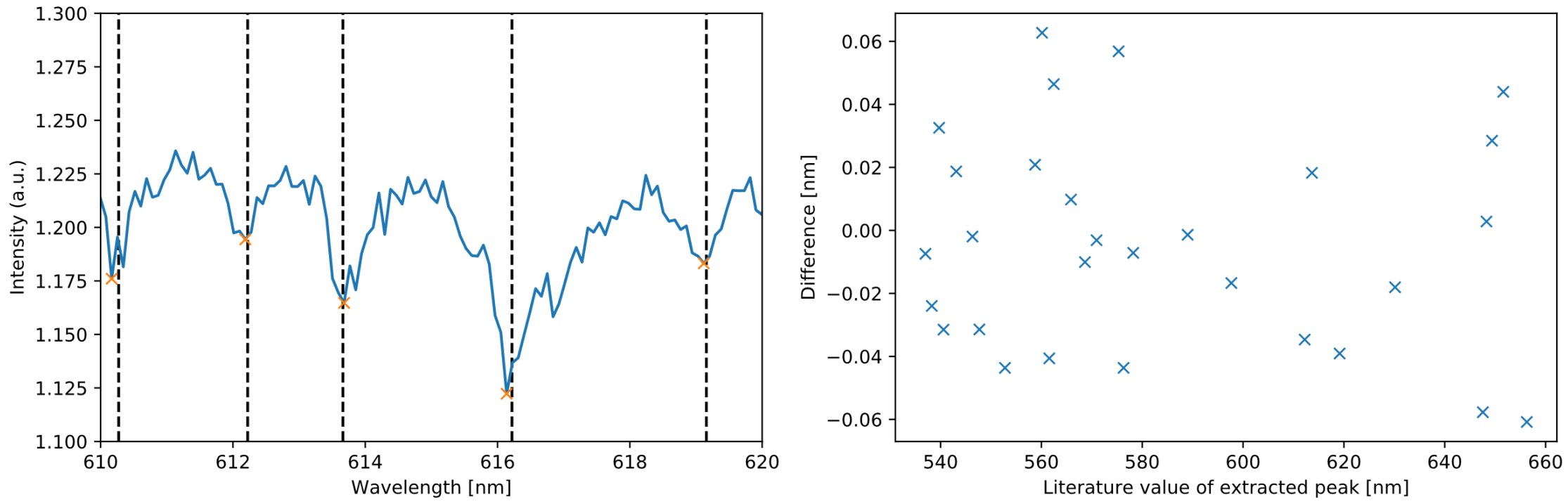

**Figure 4.** Left: Matching of literature absorption lines and detected peaks. Dashed lines are the strong absorption lines derived from Moore et al. Crossed marker is the detected peak. Right: Difference between the estimated peak wavelength and the literature value. Estimated wavelengths were calculated with Ne lamp calibration.

Table 1 Extracted Fraunhofer lines used for the calibration.

| **Calculated peak** | **Literature value** | **Difference [nm]** | **Assignments** |
|---|---|---|---|

| wavelength [nm] | [nm] | | |
|---|---|---|---|
| 536.98996247 | 536.9974 | -0.00743753 | Fe I |
| 538.23492172 | 538.2589 | -0.02397828 | Blend of Mn I (537.7614 nm), Fe I (537.9581 nm), C I (538.0322 nm), not identified (538.0737 nm), Ti II or Fe I (538.1028 nm), not identified (538.2277 nm), Fe I (538.3380 nm), Fe I (538.6340 nm), Fe I - Cr I (538.6971 nm), Fe I (538.7484 nm), (538.8351 nm), Fe I (538.9486 nm), Fe I (539.1465 nm), Fe I (539.3176 nm) |
| 539.74666731 | 539.7141 | 0.03256731 | Fe I (Ti I) |
| 540.54700232 | 540.5785 | -0.03149768 | Fe I |
| 543.21582275 | 543.1071 | 0.01872275 | Blend of Fe I (542.9706 nm), Mn I (543.2548 nm), Fe I (543.2995 nm), Fe I (543.4534 nm) |
| 546.32694552 | 546.3289 | -0.00195448 | Fe I |
| 547.66065761 | 547.6921 | -0.03144239 | Ni I |
| 552.72804687 | 552.7717 | -0.04365313 | Blend of Fe I (552.5552 nm), Sc II (552.6821 nm), Mg I (552.8418 nm) |
| 558.77122768 | 558.7504 | 0.02082768 | Blend of Fe I (558.6771 nm), Fe I (558.7581 nm), Ni I (558.7868 nm), Ca I (558.8764 nm) |
| 560.19271435 | 560.1300 | 0.06271435 | Blend of Fe I (559.8305 nm), Ca I (559.8491 nm), Ni I (560.0028 nm), Fe I (560.0234 nm), Ca I (560.1286 nm), Cr I (560.1820 nm), Ca I (560.2864 nm), not identified (560.3771 nm), Fe I (560.7669 nm), Fe I (561.5658 nm) |
| 561.52518166 | 561.5658 | -0.04061834 | Fe I |
| 562.50220937 | 562.4558 | 0.04640937 | Fe I - V I |
| 565.87658315 | 565.8668 | 0.00978135 | Fe I |
| 568.62833576 | 568.6384 | -0.01006424 | Blend of Na I (568.2647 nm), Si I (568.4493 nm), Fe I (568.6540 nm), Na I (568.8217 nm) |
| 570.93547057 | 570.9386 | -0.00312943 | Fe I |
| 575.37003189 | 575.3132 | 0.05683189 | Fe I |
| 576.25656322 | 576.3002 | -0.04363678 | Fe I |
| 578.20645411 | 578.2136 | -0.00714589 | Cu I |
| 589.00609751 | 589.0075 | -0.00140249 | Na I ($D_2$) (slightly shifted because of |
| | | | the side robe of Na I $(D_1)$) |
| 597.66197902 | 597.6787 | -0.01672098 | Fe I |
| 612.18792400 | 612.2226 | -0.03467600 | Ca I |
| 613.68066633 | 613.6624 | 0.01826633 | Fe I |
| 619.11801460 | 619.1571 | -0.0390854 | Fe I |
| 630.13274696 | 630.1508 | -0.01805304 | Fe I |
| 647.50548207 | 647.5632 | -0.05771793 | Fe I |
| 648.28371419 | 648.2809 | 0.00281419 | Ni I |
| 649.40725464 | 649.3788 | 0.02845464 | Ca I |
| 651.65227915 | 651.6803 | 0.04397915 | Fe II |
| 656.21997644 | 656.2808 | -0.06082356 | Hα |

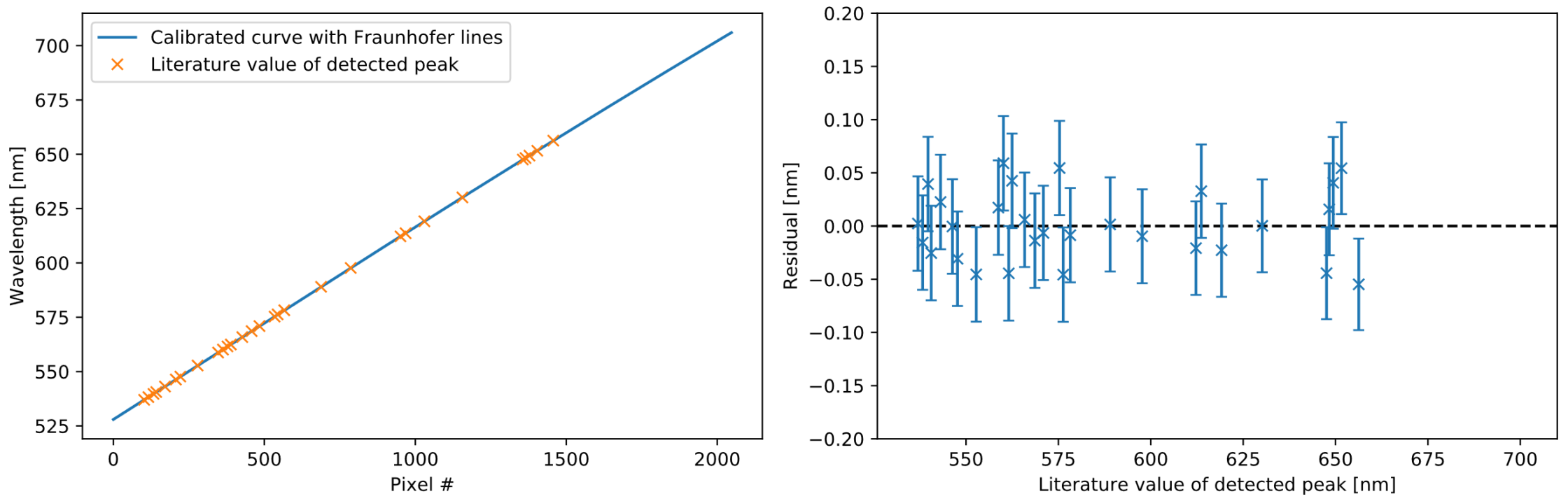

**Figure 5.** Result of the calibration with Fraunhofer lines. Left: calibrated curve and the detected peak. Right: Residual of each detected peak.

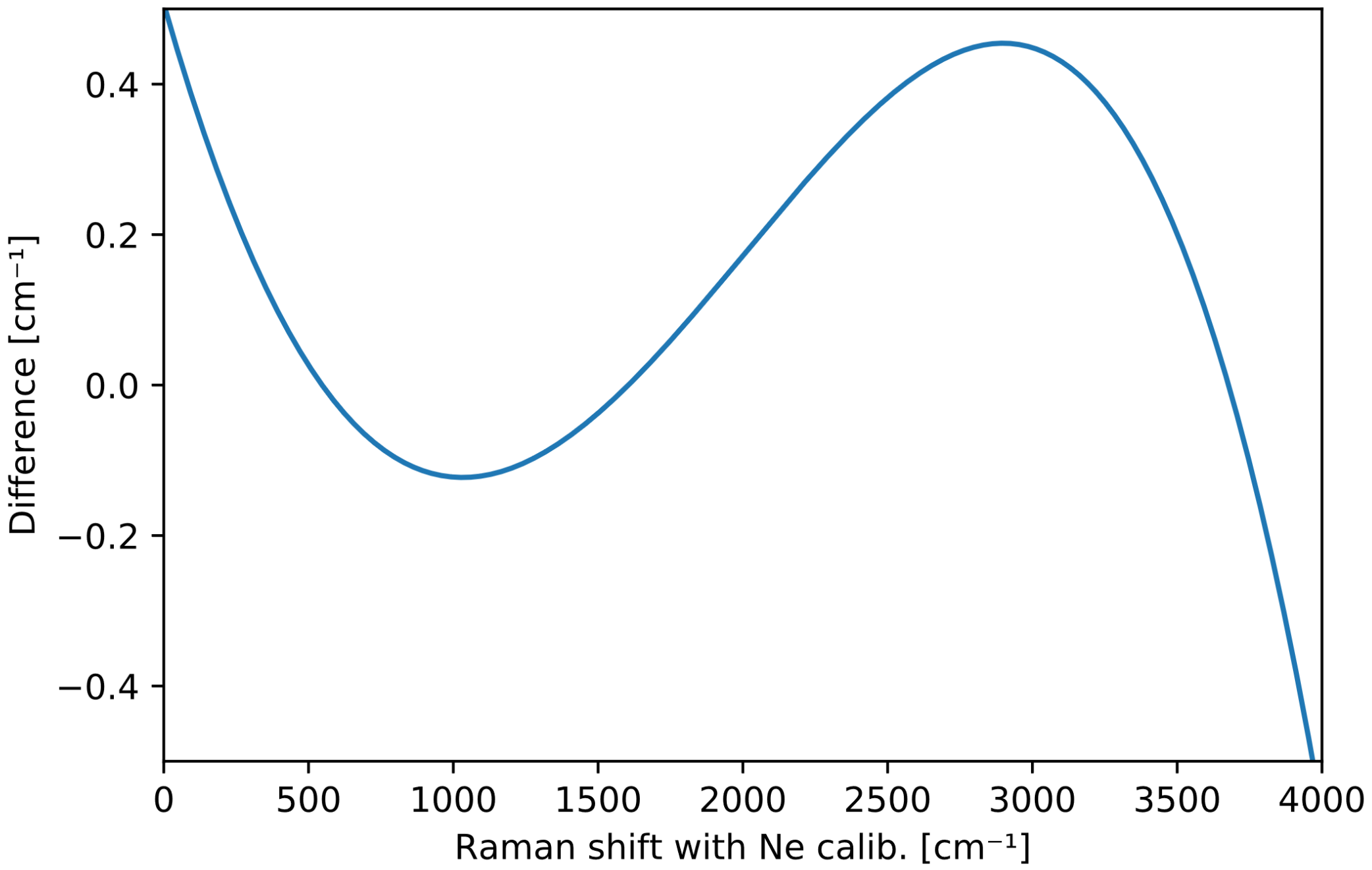

**Figure 6.** Difference of the Raman shift derived from Ne lamp calibration and Fraunhofer lines calibration.

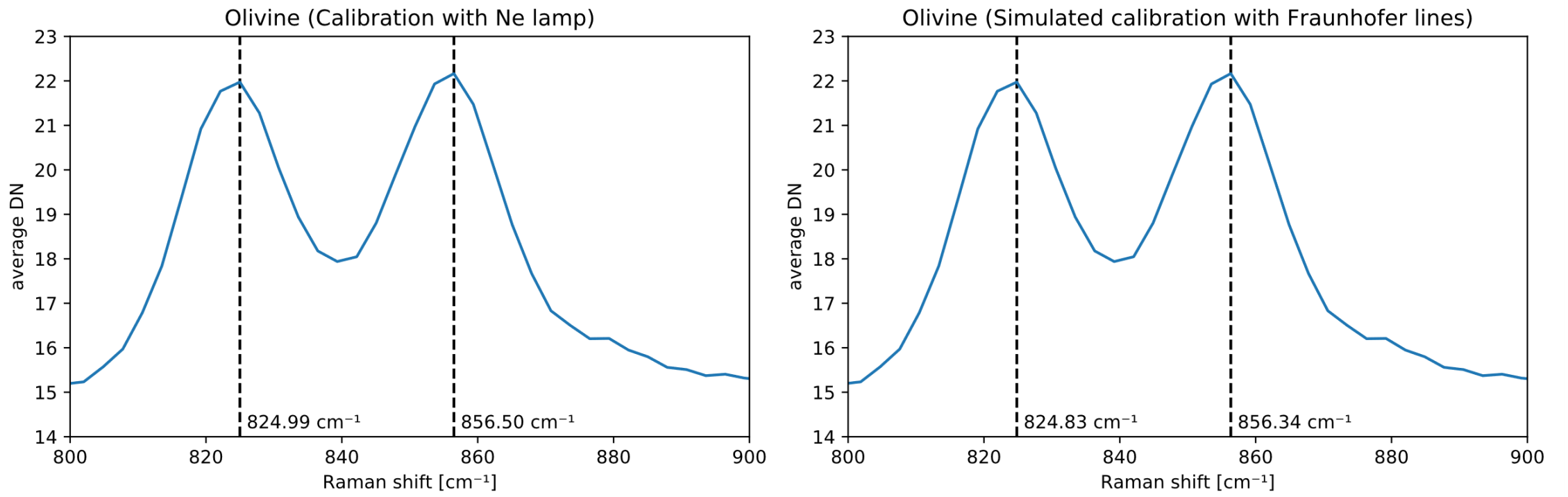

**Figure 7.** Raman spectrum doublet of olivine measured with BBM. Left: Original spectrum calibrated with neon lamp. Right: Simulated spectrum calibrated with Fraunhofer lines. Simulated spectrum is calculated by adding Figure 6 to original spectrum.

## 4. Conclusion

Possibility of utilization of Fraunhofer lines to the in-situ wavelength calibration was shown. For accurate calibration with Fraunhofer lines, determination of the device function of the single spectral line and calculation of the blended spectral peaks are important. Using this method, spectrometers for planetary exploration without calibration targets like RAX will be able to execute calibration in some conditions and to corroborate the results of the testing campaigns conducted on the ground. Thus, this is helpful for future spectrometer development for deeper space exploration because it will reduce mass and volume by removing a mechanism for looking at calibration targets.

## 5. Acknowledgments